\documentclass[aps,pre,twocolumn,floatfix,groupedaddress]{revtex4}

\usepackage{graphicx}
\usepackage{dcolumn}
\usepackage{color}
\usepackage{hyperref}
\usepackage{latexsym}
\usepackage{amsmath, amsthm, amssymb}
\usepackage{epsfig}
\usepackage{latexsym}
\usepackage{bm}
\begin{document}

\title{Co-action provides rational basis for the evolutionary success of
Pavlovian strategies}
\author{V. Sasidevan}
\email{sasidevan@imsc.res.in}
\affiliation{The Institute of Mathematical Sciences, CIT Campus - Taramani, Chennai 600113, India}
\author{Sitabhra Sinha}
\email{sitabhra@imsc.res.in}
\affiliation{The Institute of Mathematical Sciences, CIT Campus - Taramani, Chennai 600113, India}

\date{\today}


\begin{abstract}
Strategies incorporating direct reciprocity, e.g., Tit-for-Tat and
Pavlov, have been shown to be successful for playing the Iterated
Prisoners Dilemma (IPD), a paradigmatic problem for studying the
evolution of cooperation among non-kin individuals. However it is an
open question whether such reciprocal strategies can emerge as the
rational outcome of repeated interactions between selfish agents. Here
we show that adopting a co-action perspective, which takes into
account the symmetry between agents - a relevant consideration in
biological and social contexts - naturally leads to such a strategy.
For a 2-player IPD, we show that the co-action solution corresponds to
the Pavlov strategy, thereby providing a rational basis for it. For an
IPD involving many players, an instance of the Public Goods game where
cooperation is generally considered to be harder to achieve, we show
that the cooperators always outnumber defectors in the co-action
equilibrium. This can be seen as a generalization of Pavlov to
contests involving many players. In general, repeated interactions
allow rational agents to become aware of the inherent symmetry of
their situation, enabling them to achieve robust cooperation through
co-action strategies - which, in the case of IPD, is a reciprocal
Pavlovian one.
\end{abstract}

\maketitle
%
%

\section*{Introduction}
\label{sec1}
Understanding how cooperation can emerge 
in a society, each of whose individual members
are seeking to maximize their personal well-being, is one of the
fundamental problems in evolutionary biology 
and social sciences~\cite{axelrod,nowak2012,pennisi2005}. 
The ever present temptation to not cooperate
(thereby avoiding the cost associated with such an action) while
enjoying the benefits of the cooperative acts of others appears to make
cooperation unstable even if it arises by chance.
Yet cooperation is seen to occur widely in nature, ranging from
communities of 
micro-organisms~\cite{Crespi2001,Velicer2009}, cellular
aggregates~\cite{Axelrod2006}
and synthetic ecologies~\cite{Shou2007} to primate
societies~\cite{deWaal2000}.
The fragility of cooperation among unrelated individuals (i.e.,
non-kin) has been conceptually formalized in terms of 
the Prisoner's Dilemma (PD) game~\cite{Rapoport1965,Kuhn2014}
which demonstrates how the pursuit of maximal individual benefit could lead to a
collective outcome that is disastrous for all. 
Extensive investigation of this model system has revealed that
in the Iterated PD (IPD), where
repeated interactions are allowed between the same pair of
individuals, successful strategies typically use information about
previous interactions to choose the current action~\cite{Sigmund2010}.
In other words, these strategies embody the
phenomenon of direct
reciprocity that can lead to the evolution and
maintenance of cooperation~\cite{Trivers1971}. 
Empirical evidence from experiments with human and
animal subjects have been put forward in support of this notion that
cooperative behavior towards other
individuals is conditioned on the past actions of
agents~\cite{Milinski1987,Milinski1998,Hauser2003,Bshary2008,Krams2008,StPierre2009,Carter2013,Carter2015,Brandl2015}.

One of the most well-known strategies incorporating direct reciprocity
is tit-for-tat (TFT) where each agent initially cooperates and then
imitates the preceding action of its opponent in all
subsequent rounds~\cite{axelrod}. This deceptively simple strategy has
been shown to be successful in computer tournaments where different
strategies compete in playing IPD with each other.
However, it is known that TFT is vulnerable to noise arising out of
misunderstanding of intent, errors in perception and mistakes in
implementing their actions by the players - situations that arise in
most natural situations.
In such noisy environments, robust cooperation can result from other strategies 
such as generous tit-for-tat (GTFT) 
which forgives a defection with a small
probability, contrite tit-for-tat (CTFT) that follows an unintentional
defection with unconditional cooperation and
Pavlov which repeats its prior move if it has been rewarded with a
sufficiently high payoff
but changes its behavior on being punished with a low pay-off.

While these ingenious behavioral rules
have been highly effective in tournaments
where they compete in playing IPD against a variety of other
strategies, it is unclear how one would have arrived at them as the
solution to a rational decision problem. An {\em ab initio} derivation
of such a strategy incorporating direct reciprocity 
as a rational solution to IPD will not only
provide a theoretical breakthrough but may unveil new tools for
addressing different strategic interaction problems.
In this paper we show that considering the symmetry between
the players - a
relevant consideration in biological and social contexts - by using a
co-action perspective~\cite{sasi}, allows us to obtain a rational
solution for IPD, which we show to be same as Pavlov.
Even when uncertainty about the actions of other
agents and errors in implementing strategies causes
agents to defect occasionally, the information about earlier
moves can help the agents in inferring the underlying symmetry of the
situation and thereby restore cooperation, which provides a fresh
perspective on Pavlovian strategies. More importantly,
we generalize this approach to the case of IPD with many players
to address the question of cooperation in Public Goods dilemmas where
cooperation is generally considered to be harder to
achieve~\cite{Dawes1980}.
Surprisingly, we find that the cooperators always outnumber the
defectors in the co-action equilibrium of the $N$-person IPD - which
we propose as the extension of Pavlov to multi-player games.
Our results show that co-action provides a general framework to
understand why it is rational to cooperate even when it is lucrative
to act otherwise.

In the next section we first present a summary of the co-action solution
concept for single-stage (or one-shot) strategic
interactions~\cite{sasi} that is the appropriate framework for
analyzing games under conditions of complete symmetry.
As the knowledge of such symmetry may itself become apparent to agents
through repeated interactions, we next extend the co-action principle
to an iterative setting. In the following section we report the
results of solving the IPD in the co-action framework both for the
2-player as well as the $N$-player scenarios. We conclude with a brief
discussion of the implications of our results for the evolution of
cooperation and related issues.

\section*{The co-action solution concept}
\label{sec0}
The conventional Nash solution of a game defines a set of strategy choices by
agents such that no one gains by unilateral deviation, i.e., altering
only her strategy while assuming that those of others remain fixed. 
However, as we have shown recently 
for a {\em single-stage} game~\cite{sasi}, the assumptions underlying
the Nash framework are mutually inconsistent when the game situation
is symmetric (i.e., exchanging the identities of the agents leaves the
payoff structure invariant).  
Specifically, assuming that (a) each agent is aware that all
agents are equally capable of analyzing the game situation,
is inconsistent with the assumption that (b) agents can make
unilateral deviations
in their strategy - a necessary premise for 
obtaining a dominant strategy.
Removing this inconsistency yields the
co-action equilibrium\cite{sasi,sasidevan} where 
each agent is aware of the symmetric situation that all agents are in.
Thus an agent will realize that, whatever strategy choice she
is going to make, 
other agents, being in the
same symmetric situation and being just as rational as her, will make
too. 
Simply put, this the only logical conclusion that can be arrived at by
a rational agent. Note that this does not imply that agents will
necessarily choose the same action, e.g., if they are using mixed
strategies as may happen in PD for a certain range of payoff values 
as discussed below.

\begin{figure}
\centering
 \includegraphics[scale = .30]{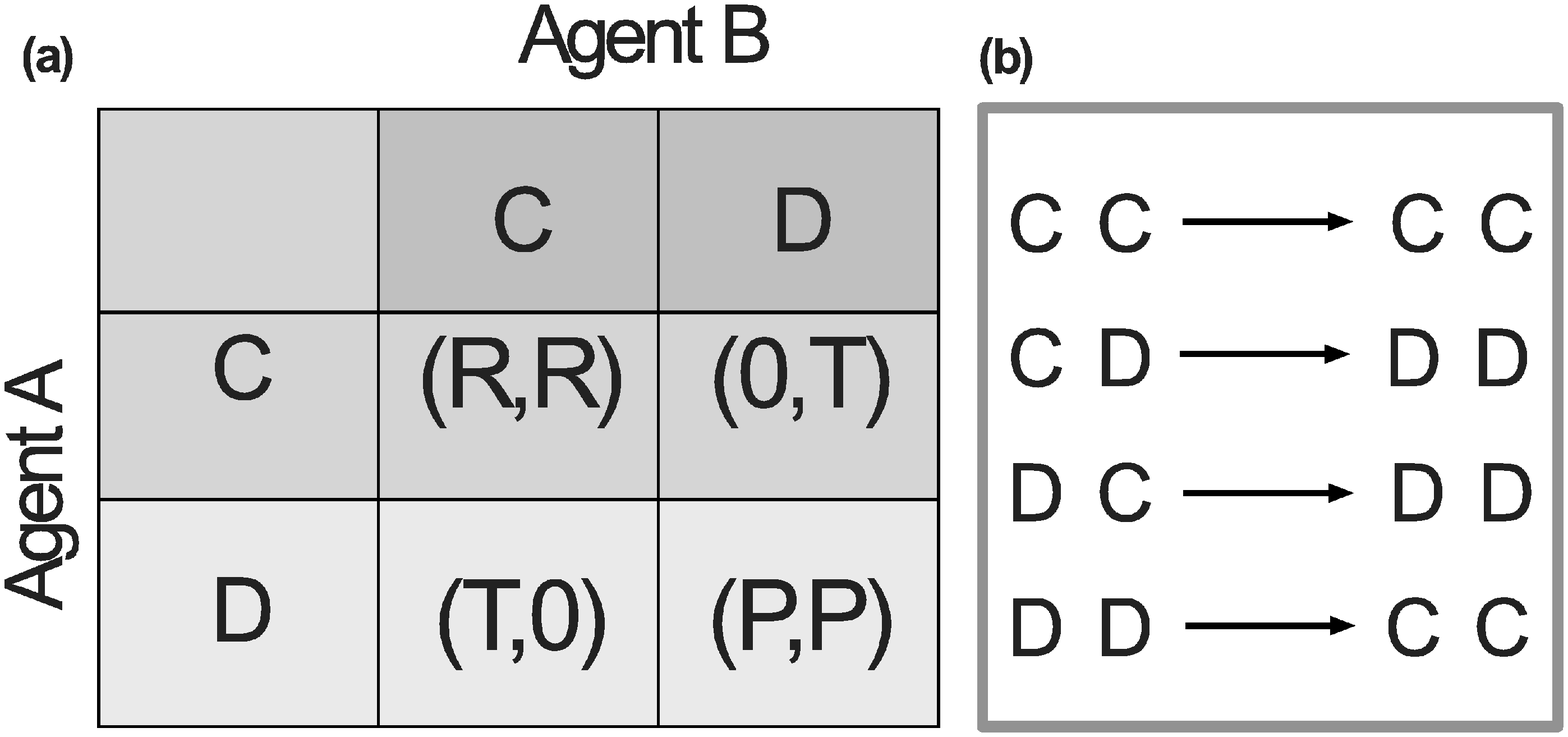}
 \caption{(a) A generic representation of the payoff matrix for a
two-person PD ($T>R>P>0$, with the sucker's payoff $S$ assumed to be zero
for convenience). At every round, each agent can choose one of two
possible actions, cooperate (C) or defect (D). For each pair of
actions, the first entry in each payoff pair belongs
to Agent A while the second belongs to Agent B. \\
(b) Representation of the Pavlov strategy
corresponding to the co-action solution of the two-agent IPD. The
arrows connect the optimal actions of Agents A and B (in order) in the present
round to information about their actions in the previous round.}
\label{fig1}
\end{figure}

We illustrate this distinction between the Nash and co-action
frameworks in the single-stage PD representing
a one-off strategic encounter between two
agents who have the choice to either cooperate ($C$) or defect ($D$).
In this game, if
both agents choose $C$, each receives a reward payoff $R$, while if they
both choose $D$, each is penalized with a punishment payoff $P<R$.
If they choose different actions, then the defector receives the
highest payoff
$T >R$ (the temptation to defect) while the cooperator gets the lowest
(or sucker's) payoff $S <P$ [see the payoff matrix in Fig.~\ref{fig1}~(a)]. 
Thus, the payoffs are ordered as $T>R>P>S$, for which it is easy to
see that mutual defection is the only Nash equilibrium.
In contrast to the Nash solution, co-action leads to a ``cooperative''
outcome resulting from the agents maximizing their payoffs under the
assumption that other agents will use the same strategy as them
(although they may not
necessarily choose the same action - $C$ or $D$ - if the strategy is a
probabilistic one). In the single-stage PD, this amounts to
determining that value of $p$, i.e., the probability that an agent
will choose $C$, for
which the expected payoff function for each of the agents
$W = p^2 (R + P - T - S) + p (T + S - 2P ) + P$ is maximum.
Note that, here we have used the key concept of co-action, viz., that
each agent will independently choose $C$ with the same probability $p$.
It is easy to see that this optimization problem has a unique solution
corresponding to the agents always cooperating when $T \leq 2R$ and
cooperating with a probability $p^* = (2P-T)/[2(R+P-T)]$ when $T >
2R$~\cite{sasi}.

The above argument does not take into account the possibility of 
previous interactions among the agents. In other words, there is no
consideration of any memory of how the agents behaved in previous
rounds. However, many strategic interactions that arise in
biological, economic and social contexts are iterative in nature, 
where individuals can engage with each other repeatedly. 
If the agents are capable of recalling how their
opponents acted in earlier interactions, this information can be used
by them to formulate their current strategies. The phenomenon of
direct reciprocity~\cite{Sigmund2010} can be placed in this general
context, providing a platform for addressing the problem of evolution
of cooperation through the Iterated
Prisoners Dilemma (IPD).
In contrast to the single stage game described above, the
IPD involves two players repeatedly playing the
game. 
Just as for the single stage
game, mutual defection is the only Nash equilibrium  for a finitely
repeated IPD, which can be easily shown by a backward
induction argument. In an infinitely repeated IPD however, it is possible
to have mutual cooperation as an equilibrium outcome, as indicated by folk
theorems~\cite{fundenberg}.
Computerized tournaments in which different programs are made to play
IPD
against each other have indeed shown the success of strategies that
incorporate reciprocity, such as TFT and
Pavlov, which can help maintain cooperation~\cite{axelrod}. It would be a
significant theoretical breakthrough if any of these reciprocal
strategies can be shown to be the rational solution of IPD - even in a
restricted context such as that of 1-step memory rules (i.e., those which
take into account the history of only the previous round). As we show
below, this can be achieved using the co-action solution concept.
Note that direct reciprocity allows the knowledge gained from
previous interactions to be used by agents to infer the existence of
symmetry - even in the absence of any communication
between them - which is the crucial ingredient for the co-action concept to
apply. Thus, generalizing the co-action framework which had been
originally proposed in the context of one-shot games to an iterative setting
allows its application to a wide class of non-cooperative strategic
interactions in nature
where symmetry between the players need not be
assumed a priori.

\section*{Results}
\label{sec2}
\subsection*{IPD between two agents}
The co-action solution for the case of two agents playing IPD can be
derived as follows. Consider the 
payoff matrix for a single round of interaction between the agents as
shown in Fig.~\ref{fig1}~(a). The
value of the lowest payoff $S$ is assumed to be zero without loss of
generality.
In addition, we consider the case $2R>T$ so as to rule out the
possibility of a
strategy in which agents take turns to alternately cooperate and
defect.
For the sake of clarity, we look at 1-step memory strategies
where each agent has the knowledge of the choice made by all
agents in the last round.
Similar considerations will apply when
extending the analysis to longer-memory strategies.

In the co-action equilibrium, the symmetry of the game situation as
perceived by the agents governs their strategies. As the agents can
recall their actions in the immediately preceding round of the game,
if both had chosen the same action (i.e., $CC$ or $DD$), this is recognized as
establishing complete symmetry between the agents - in which case,
they behave as in the single-stage PD co-action solution~\cite{sasi}.
If, on the other hand, each had chosen a different action (i.e., $CD$ or
$DC$), then the agents realize that they are not in a symmetric
situation and will resort to Nash-like reasoning.

To set out the argument in detail, we consider the four different
possibilities that can arise during the course of the IPD, viz., (i)
agent $A$ cooperated while agent $B$ defected ($CD$),
(ii) both cooperated ($CC$), (iii) both defected ($DD$) and 
(iv) agent $A$ defected while agent $B$ cooperated ($DC$), in the last
round. 
Thus, the state an agent is in at any given time could be any one of the
following: $|1\rangle = |C,1\rangle$, $|2\rangle
=|C,2\rangle$, $|3\rangle =|D,0\rangle$ and $|4\rangle =|D,1\rangle$.
In this notation, the first entry denotes whether the agent cooperated ($C$) or
defected ($D$) and the second entry denotes the
total number of agents who cooperated in the previous round.  If 
$p_i$ denotes the probability with which an agent in state $|i\rangle$ 
switches her action, we can express her expected 
payoffs $W_i$ in the different states as:
\begin{align}
\label{eqn1}
W_1 &= p_1\left(P + p_4\left(T-P-R\right)\right) + p_4 R,\\
\label{eqn2}
W_2 &= R-p_2 (2 R-T)-p_2^2 (T -P - R ),\\
\label{eqn3}
W_3 &= P + p_3 (T-2P) +p^{2}_3 (R+P-T),\\
\label{eqn4}
W_4 &= T - p_4\left(T-R-p_1\left(T-R-P\right)\right) -
p_1\left(T-P\right).
\end{align}
Note that the payoff  $W_2$ is a function of only $p_2$ and $W_3$ is a
function of only $p_3$, as, in the co-action framework, 
both the agents in these states (corresponding to $CC$ and $DD$, 
respectively) are in a completely
symmetric situation. Hence, the agents in state $|2\rangle$ ($|3\rangle$)
will each switch to defection (cooperation) with the same probability $p_2$
($p_3$). It is easily seen that the values for $p_2$ and $p_3$
that maximize the respective payoff functions $W_2$ and $W_3$
are $0$ and $1$, respectively (corresponding to mutual cooperation).

For the states $|1\rangle$ and $|4\rangle$ (corresponding to $CD$ and
$DC$, respectively), where the agents are not in a symmetric
situation, agent in state $|1\rangle$ will try to maximize $W_1$ by
varying $p_1$ for any given value of $p_4$ while the agent in state
$|4\rangle$ will seek to maximize $W_4$ by varying $p_4$ for any given
value of $p_1$. Using the same reasoning that is employed to obtain
the Nash strategies, it is easy to see that the only mutually consistent
choice for the optimal strategies of the two agents is $p_1^* = 1$ and $p_4^*
= 0$ (corresponding to mutual defection). 
The optimal strategies for the agents in different states are 
summarized in Fig.~\ref{fig1}~(b).
Hence, the agents will resort to co-action thinking whenever they find
themselves in a symmetric situation (as in $CC$ or $DD$) while they
use Nash-like reasoning in other situations (as in $CD$ or $DC$). 
In the latter case, they will arrive at a symmetric situation
in the next round (as they choose $DD$), and thereafter will mutually
cooperate.

An important observation about the co-action solution of the two-person
IPD discussed above is that the optimal strategy [Fig.~\ref{fig1}~(b)]
turns out to be the same as the Pavlov strategy for IPD
proposed by Nowak and Sigmund~\cite{nowak}. This strategy has been
shown to have certain advantages over the well-known tit-for-tat (TFT)
strategy~\cite{axelrod} for playing IPD, viz., it can correct for
occasional mistakes in implementation of strategies and
can exploit unconditional cooperators~\cite{Kraines1989,Imhof2007}.
More generally, Pavlov
type of behavior has been widely observed in natural
situations~\cite{domjan}, including experimental realizations of
PD~\cite{wedekind}.
We emphasize that unlike in earlier studies where the
Pavlov strategy is considered as an {\em ad-hoc} behavioral rule for
agents, here we have demonstrated from first principles that
such a strategy is the optimal solution for
rational, selfish agents in the
two-agent IPD.

\subsection*{IPD between many agents}
\label{sec3}
We now consider an IPD with $N (>2)$ agents, each of whom play with
all the others in every round. An individual chooses an action (either
$C$ or $D$) in each round which it employs against everyone else in that
round, receiving payoffs for each pairwise interaction according to
the matrix in
Fig.~\ref{fig1}~(a). As in the two-agent case, we assume that $S=0$ and
$2R > T$. In addition, we set the ``punishment'' payoff $P$ to 0 for
simplicity (alternatively, one can consider 
$P = \epsilon \ll 1$~\cite{nowak_2}). The total payoff received by an
agent in any round is the sum of the individual payoffs from each of
the $(N-1)$ two-agent interactions.
This ensures that all the agents receive a lower payoff if
everyone defects than if they all cooperate,
and if any agent switches from $D$ to $C$, the average payoff of the
agents increases.

The above situation describes an instance of {\em public goods dilemmas} where individual
contributions towards a public good increases the collective benefit
although the cost borne by an individual for this contribution exceeds
the benefit she derives from it~\cite{Davis1993,Kuhn2014}.
While the general problem of public goods has been considered under
various guises in the literature~\cite{Olson1965}, in the simple
quantitative setting involving a well-mixed population as described above,
it is easy to see that a single round of interaction in a $N$-person
public goods game is equivalent to $N-1$ pairwise PD
interactions~\cite{Hauert2003}.
This does not imply that the situation described by the public goods
dilemma simply corresponds to a
quantitative increase in the number of agents of the PD game,
but rather
involves a profound change in the nature of the
interactions~\cite{Kollock1998}. Agents can react only to the combined
effect of the actions of all other agents and not to the individual
strategies of specific agents. The anonymity provided
to individuals in the multi-player setting means that they are more
likely to defect (i.e., act as free-riders) without much fear of retaliation 
by others~\cite{Dawes1980}. 
	
The state that an agent is in at any given time can be represented by
either $|C,n\rangle$ or $|D,n\rangle$ according to 
whether she cooperated ($C$) or defected ($D$) in the
previous round, with $n$ denoting the total number of agents who
cooperated in the previous round. 
In the co-action framework, the set of agents who played $C$
in a particular round realizes that all of them who chose $C$ 
are in a symmetric situation. Similarly, the set of agents who played
$D$ are aware of the symmetry among them. Thus, within each group,
all agents will use identical strategies for the next round.
For simplicity, we consider only pure strategies where agents choose
either $C$ or $D$ with probability 1~\cite{Dawes1980,hauert,boyd,axelrod_2}.

Let us first consider the two extreme cases corresponding to either
everyone cooperating or everyone defecting in the previous round. 
If all the agents had cooperated, they would realize that all of them
would use
identical strategies. The expected payoff of any agent is simply an
integral multiple of $W_2$ (see Eq.~\ref{eqn2}), the corresponding payoff in the two-agent
case studied earlier. Thus, on optimizing payoff, all agents
choose $C$ in the next round. By similar arguments, if all agents
had chosen to defect in the previous round, they would choose $C$
in the next round. 

\begin{figure}
\centering
\includegraphics[scale = .30]{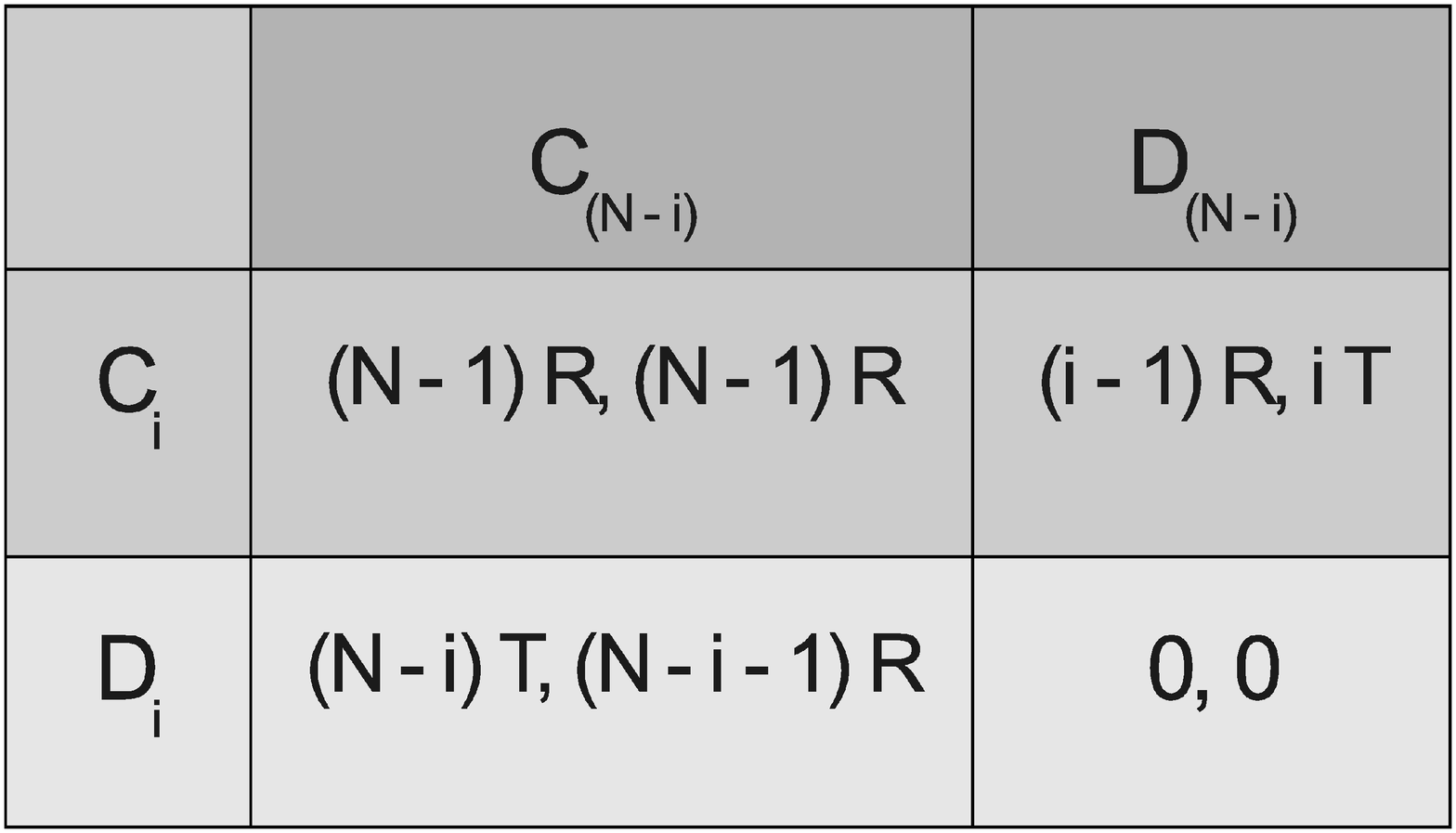}
\caption{ Representation of possible payoffs in an $N$-player IPD
under the co-action framework, when $i$ agents cooperated and $N-i$ agents 
defected in the previous round.}
\label{fig3}
\end{figure}

When some of the $N$ agents cooperate and the others defect, we can
treat the situation as identical to a two-player scenario, 
with the Nash equilibrium
being the optimal strategy. Note however that each ``player'' is
now a group of agents and the corresponding Nash solution is
distinct from the one corresponding to everyone defecting as is
obtained in a conventional 2-person PD.
The expected payoffs of the two sets of agents can be conveniently
represented by a two-player payoff matrix as shown in Fig.~\ref{fig3}.
Here the row corresponds to the set of $i$ agents (where $i =
1,2,\ldots,N-1$) who cooperated in
the last round, while the column corresponds to the set of $(N-i)$
agents who defected. In the next round, the row ``player'' can either
choose to continue cooperating ($C_i$) or switch to defection ($D_i$).
Similarly, the column ``player'' can switch to cooperation in the next
round ($C_{N-i}$) or continue to defect ($D_{N-i}$).
Thus, starting with any combination of cooperating and defecting
agents, we can obtain the optimal strategies for the two sets of
agents which depend on the ratio of the payoffs $T/R$
for a given $i$.

The co-action solution for the four possible situations that can
arise in terms of the relative magnitudes of the payoffs for the two
sets of agents are:
\begin{itemize}
\item $(N-1)R \geq iT$ and $(N-1)R \geq (N-i)T$: From Fig.~\ref{fig3},
it is clear that cooperation is the optimal choice for both the
sets of agents as neither will benefit by deviating from this strategy.\\ 
\item $(N-1)R \geq iT$ and $(N-1)R < (N-i)T$: It is easy to see that
cooperation is the optimal choice for the column ``player''
independent of the action of the row ``player'', and using this
information, one observes that the optimal choice for the row
``player'' would be to defect.
Thus, the set of agents who cooperated in the
previous round will switch to defection, while the set which defected
will switch to cooperation.\\
\item $(N-1)R < iT$ and $(N-1)R \geq (N-i)T$: Again it is easy to see
that cooperation is the optimal choice for the row ``player'' 
independent of the action of the column ``player'', and using this 
information, one observes that the optimal choice for the column
``player'' would be to defect. The agents will therefore continue with
the same actions as in the previous round.\\
\item $(N-1)R < iT$ and $(N-1)R < (N-i)T$: This situation arises
only when $i = N/2$ (and hence only for even values of $N$), i.e.,
when there are equal number of cooperators and defectors. For this
case, there are two possibilities for the optimal action, one where
the ``players'' continue with the same action as in the previous
round, and, the other where each of them switches to the opposite
action. Note that the level of cooperation does not change in either
of the cases.
\end{itemize}

For illustrative purpose, we now discuss in detail the co-action
solution of the
$N$-person IPD for the cases when $N = 3$, $4$ and $5$. In each of
these cases, we shall denote the distinct states that are possible for
the system to be in as $S_j$ where $j=0,\ldots, N$ is the number of
cooperators in that state.
For $N = 3$ agents, it is easy to see by referring to the general
co-action solution given above that
the optimal strategies will result in the
following evolution between the distinct states of the system:
$S_0 \rightarrow S_3$, $S_1 \rightarrow S_2$, $S_2 \rightarrow S_2$
and $S_3 \rightarrow S_3$.
Thus, if all three agents had chosen the same action ($C$ or $D$) in the
previous round, all of them cooperate in the next round ($S_3$). 
In all the other cases, the system converges to the state $S_2$
corresponding to two cooperators and one defector.
This result clearly distinguishes the co-action approach from the
conventional Nash solution, which would have corresponded to all three
defecting. A notable feature of the co-action solution is the
stable coexistence of cooperators and defectors (as in state $S_2$).

For the case when $N=4$, as before by referring to the general
co-action solution above, we see that the optimal strategies will result in the
following evolution between the distinct states of the system:
$S_0 \rightarrow S_4$, $S_1 \rightarrow S_3$, $S_2 \rightarrow S_4$
(if $3R \geq 2T$) or $S_2 \rightarrow S_2$ (otherwise),
$S_3 \rightarrow S_3$ and $S_4 \rightarrow S_4$.
We can see that for $N = 4$ (unlike for $N = 2$ and $3$) the
solution begins to depend on the ratio of $T$ to $R$, which is also true
for all higher values of $N$. 

As a final example, we consider the case when $N=5$.
Here the optimal strategies depend on whether the magnitude of the
payoff values satisfy $4R>3T$. If this is true,
it will result in the
following evolution between the distinct states of the system:
$S_0 \rightarrow S_5$, $S_1 \rightarrow S_4$, $S_2 \rightarrow S_5$,
$S_3 \rightarrow S_5$, $S_4 \rightarrow S_4$ and $S_5 \rightarrow S_5.$
On the other hand, if $4R<3T$, the following
evolution will be observed:
$S_0 \rightarrow S_5$, $S_1 \rightarrow S_4$, $S_2 \rightarrow S_3$,
$S_3 \rightarrow S_3$, $S_4 \rightarrow S_4$ and $S_5 \rightarrow S_5$.

Thus, we can draw the following general conclusions:
(a) the state in which everybody cooperates (i.e., $i = N$) is a stable
state, (b) a state in which all but one agent cooperate ($i = N-1$)
is also a stable state, (c) states where the defectors are in a minority
are stable if $T/R > (N-1)/(N-i)$ and (d) when the cooperators
are in a minority, in the next iteration all agents will cooperate
if $T/R < (N-1)/(N-i)$, otherwise they will switch their
respective choices (from C to D and vice versa). In the special
case when $N$ is even with exactly half of the agents cooperating and
$T/R > 2(N-1)/(N)$, multiple equilibria are possible.
The most important point to note from the above results is that
cooperators can coexist with defectors, and moreover, always form a
majority, in the co-action solution of the $N$-player IPD.

\section*{Discussion}
\label{sec4}
In contrast to the conventional wisdom that
defection should be the preferred strategy of selfish agents,
human subjects playing PD in either single-stage
or multiple round experiments do achieve some
measure of cooperation (e.g., see Ref.~\cite{Sally1995}).  
Understanding how such cooperation arises can be
investigated in the context of repeated interactions between
agents. In this case, agents can ``remember'' their past actions and
the resulting outcomes, and they can use this information to govern 
their future decisions - a phenomenon referred to as direct
reciprocity~\cite{Trivers1971}. 
Apart from this, several other mechanisms for the emergence of
cooperation through natural selection
have been proposed~\cite{Nowak2006}, such as, kin
selection~\cite{Hamilton1964}, indirect
reciprocity~\cite{Nowak1998}, network
reciprocity~\cite{nowak_2,Lieberman2005} and group
selection~\cite{Traulsen2006}. 
Even within the conventional game theoretic framework, there have been
formal attempts at modifying IPD so as to make cooperation viable,
involving concepts such as $\epsilon$-equilibria~\cite{radner}, incomplete
information~\cite{kreps}, bounded rationality~\cite{neyman_2}, absence of
common knowledge about the rationality of players~\cite{pettit} and
the number of iterations~\cite{neyman_1}, etc. 
%
In recent times there has been an increased focus on the evolution
of cooperation in spatially extended situations where agents
interact with multiple neighbors defined by an underlying connection
topology~\cite{nowak_2,Perc2008,Szolnoki2012,Szolnoki2012b,Szolnoki2016}.

In this paper we have addressed the question of
whether a strategy incorporating direct reciprocity that allows for
cooperation to be maintained can emerge as a rational solution of IPD.
The novel perspective that we bring to bear involves recognizing the
symmetry between agents - a crucial ingredient for the co-action
framework to apply. In an iterative setting, agents
become aware of their symmetry with other agents through the
knowledge of their actions in previous encounters. 
The most important result of our study is that cooperators and
defectors coexist in the co-action solution of the $N$-player IPD; 
moreover, the majority of
agents cooperate. This is remarkable in view of the conventional
wisdom that cooperation is extremely difficult to achieve in a group
of {\em selfish rational} agents~\cite{axelrod}. 
For the case of two players, the co-action
solution of IPD corresponds to the well-known Pavlov
strategy that has been attested in animal behavior and social
interactions~\cite{nowak}. To the best of our knowledge, the approach
we present here is the only one
that provides a rational game-theoretic basis
for such a strategy, as opposed to proposing it as a {\em ad hoc}
behavioral rule.
It also allows the generalization of Pavlov
to the situation of multiple ($N>2$) agents.

An important consideration in studies of IPD is the role of noise that can
arise from the incorrect implementation of intended action by
agents~\cite{Vukov2006,Szolnoki2009}.
Such noise
may also be due to the misinterpretation of actions of other
agents~\cite{Neill2001}. 
For example, the TFT strategy in IPD is vulnerable to such
noise as it cannot correct for occasional mistakes by agents.
While for the case of two players it is known that the Pavlov 
strategy (which is the co-action solution for $N=2$) is
stable in the presence of noise~\cite{Posch1999}, it is easy to see that even in the
case of $N>2$ agents, the co-action solution is not significantly
affected by intermittent errors on the part of the agents.


%
The iterative game situation considered here corresponds to 1-step
memory where the agents only retain information about the action of other
agents in the immediately preceding round.
The co-action concept can be easily extended to the more general
situation of agents with longer memories, once the key question of how
the symmetry among agents is to be defined in such situations is
addressed.
One possibility is that all agents who have cooperated an equal number of times
in the past are considered to be in a symmetric situation. They
will therefore adopt the co-action strategy in the next round. 
For two agents with finite memory, this will eventually lead to both
of them cooperating.
If there are more than two agents, the co-action principle suggests that those
who display similar propensities to cooperate - i.e., they have
cooperated an equal number of times in the past - will form a group
defined by complete symmetry among the agents comprising it. Thus, the
entire set of $N$ agents can be divided into a number of such
``symmetry groups''. This defines a novel class of strategic
interactions where the
``players'' are the different symmetry groups (each
consisting of one or more agents) playing according to strategies
given by the Nash equilibrium. It is important to point out that this
will not result in all agents resorting to defection as expected in
the conventional Nash framework.
Potentially, this new class of games can be used to analyze
multi-agent strategic interactions in many different contexts.

It is intriguing to consider the implications of the co-action
strategy for the behavior of individuals in real-world social
interactions. As we show here, the stable solutions are those where a
majority of agents cooperate, suggesting that the presence of a few
defectors will not necessarily result in the breakdown of cooperation
in a society. This is because rational agents who perceive each other to be
similar, will not be deterred from cooperating as long as they receive
enough mutual support - in the form of acts of cooperation - from similar 
agents. The co-action framework, therefore, implies that significant
levels of cooperation
will be seen in interactions among rational individuals in IPD-like
situations, in contrast to conventional wisdom.
There have been a large number of experiments carried out with human
subjects playing IPD (both the
2-person as well as the multiple-player version, viz., the repeated public
goods game). Surveying the results reported in many 
experiments over several decades reveal that, both for 
the two-person IPD~\cite{Sally1995}, as well as, the 
repeated public goods game~\cite{Andreoni1995,Ledyard1997}, the
majority of experimental subjects do not behave in the manner
predicted by conventional game theory. 
As shown in this paper, the co-action paradigm provides a mechanism
for a rational explanation of experiments on
IPD that do not show complete absence of cooperation. 
It can also perhaps help in
understanding cooperative behavior seen among non-human
animals who do not share kinship~\cite{Carter2015}, a phenomenon that
has been experimentally investigated in an IPD
framework, e.g., in birds~\cite{StPierre2009}.

%
The setting in which we have discussed the problem of evolution of
cooperation above corresponds to the idealized situation of fully
rational agents interacting with each other.
While the rationality assumption is used widely in situations 
involving human actors, one can ask how the co-action paradigm may
apply to other animals or even colonies of unicellular organisms where
the emergence of cooperative behavior is
observed~\cite{Crespi2001,Velicer2009,Axelrod2006,Shou2007,deWaal2000}.
As symmetry is the crucial ingredient for the co-action framework to
be valid, it is not unreasonable to apply it for interactions among
members of a homogeneous population
who share a common identity. 
This homogeneity could be in terms of, for example, the genetic composition,
physiognomy or even acquired traits. Depending on the specific
context in which the evolution of cooperation is being considered, one
or more of these identities could be relevant for the co-action framework
to apply. For instance, tag-based cooperation among ``similar''
individuals~\cite{Riolo2001}
could arise naturally under this framework.

To conclude, we have shown that the co-action paradigm
provides a new perspective to the evolution of cooperation.
The co-action concept has been earlier shown to resolve social dilemmas in
single-stage symmetric games. Here we show how the idea of co-action applies to
the more general setting of iterative game situations. Information
about previous interactions allows agents to infer symmetry (or its
absence) among themselves, allowing cooperation to emerge even when
agents had initially chosen to defect. 
The co-action framework also provides a rational basis for the
Pavlov strategy
that has been proposed for the two-person IPD, and generalizes such a
strategy to the case of several agents. 
In general, we observe that cooperators and defectors can coexist in
the $N$-player Iterated Prisoners Dilemma game, with the cooperators
constituting the majority. This is a surprising feature given the
conventional expectation that selfish, rational agents will always defect.

\section*{Acknowledgements}
This work was partially supported by the IMSc Econophysics project
funded by the Department of Atomic Energy, Government of India. We thank Deepak Dhar for helpful discussions and Renjan John for useful
comments on the manuscript.


\begin{thebibliography}{99}
\bibitem{axelrod} Axelrod, R. {\em The Evolution of Cooperation}
(Basic Books, New York, 1984).
\bibitem{nowak2012}  Nowak, M. \& Highfield, R  {\em Super Cooperators}
(Free Press, New York, 2011).
\bibitem{pennisi2005} Pennisi, E.  How did cooperative behavior
evolve~? {\em Science} 309:93. (2005).
\bibitem{Crespi2001}
Crespi, B. J.  The evolution of social behavior in microorganisms.
{\em Trends Ecol. Evol.} {\bf 16(4)}, 178-183 (2001).

\bibitem{Velicer2009}
Velicer, G. J. \& Vos, M.  Sociobiology of the Myxobacteria.
{\em Annu. Rev. Microbiol.} {\bf 63}, 599-623 (2009).

\bibitem{Axelrod2006}
Axelrod, R., Axelrod, D. E. \& Pienta, K. J.  Evolution of cooperation
among tumor cells. {\em Proc. Natl. Acad. Sci. USA}
{\bf 103(36)}, 13474-13479 (2006).

\bibitem{Shou2007}
Shou, W., Ram, S. \& Vilar, J. M. G.  Synthetic cooperation in engineered
yeast populations. {\em Proc. Natl. Acad. Sci. USA} {\bf 104(6)}, 1877-1882 (2007).

\bibitem{deWaal2000}
de Waal, F. B. M.  Primates - a natural heritage of conflict
resolution. {\em Science} {\bf 289}, 586-590 (2000).

\bibitem{Rapoport1965} Rapoport, A. \& Chammah, A. M. {\em Prisoners
Dilemma} (Univ of Michigan Press, Ann Arbor MI, 1965).

\bibitem{Kuhn2014}  Kuhn, S. Prisoner's Dilemma, in {\em Stanford
Encyclopedia of Philosophy} (ed. E. N. Zalta, Fall 2014 edition)
URL
http://plato.stanford.edu/archives/fall2014/entries/prisoner-dilemma/.

\bibitem{Sigmund2010} Sigmund, K.  {\em The Calculus of
Selfishness} (Princeton University Press, Princeton NJ, 2010).

\bibitem{Trivers1971} Trivers, R. L.  The evolution of reciprocal
altruism. {\em Q. Rev. Biol.} {\bf 46}, 35-57 (1971).

\bibitem{Milinski1987} Millinski, M.   Tit for tat in sticklebacks and the evolution of cooperation. {\em Nature} {\bf 325}, 433-435 (1987).

\bibitem{Milinski1998} Milinski, M. \&  Wedekind, C.  Working memory constrains human cooperation in the Prisoner’s Dilemma.
{\em Proc. Natl Acad. Sci. USA} {\bf 95(13)}, 755–758 (1998).

\bibitem{Hauser2003} Hauser,  M. D., Chen, M. K., Chen, F. \& Chuang, E.  Give unto others: genetically unrelated cotton-top
tamarin monkeys preferentially give food to those who altruistically
give food back. {\em P. Roy. Soc. Lond. B} {\bf 270}, 2363-2370 (2003).

\bibitem{Bshary2008} Bshary, R., Grutter, A. S., Willener, A. S. T. \& Leimar, O.  Pairs of cooperating cleaner fish provide better
service quality than singletons. {\em Nature} {\bf 455}, 964-967 (2008).

\bibitem{Krams2008} Krams, I., Krama, T., Igaune, K. \&  Mand, R.  Experimental evidence of reciprocal altruism in the pied
flycatcher Ficedula hypoleuca. {\em Behav. Ecol. Sociobiol.} {\bf 62}, 599-605 (2008).

\bibitem{StPierre2009} St-Pierre, A., Larose, K. \& Dubois, F.
Long-term social bonds promote cooperation in the iterated Prisoner's
Dilemma. {\em P. Roy. Soc. Lond. B} {\bf 276(1676)}, 4223-8 (2009).

\bibitem{Carter2013} Carter, G. G. \&  Wilkinson, G. S.  Food sharing
in vampire bats: reciprocal help predicts donations more than
relatedness or harassment. {\em P. Roy. Soc. Lond. B} {\bf 280(1753)}, 20122573 (2013). 
 

\bibitem{Carter2015} Carter, G. G. \&  Wilkinson, G. S.  Social
benefits of non-kin food sharing by female vampire bats. {\em P. Roy.
Soc. Lond. B} {\bf 282(1819)}, 20152524 (2015).

\bibitem{Brandl2015}  Brandl1,  S. J. \&  Bellwood, D. R.  Coordinated vigilance provides evidence for direct reciprocity in
coral reef fishes. {\em Scientific Reports} {\bf 5}, 14556 (2015).

\bibitem{sasi}  Sasidevan, V. \& Sinha, S.  Symmetry warrants rational
cooperation by co-action in social dilemmas. {\em Scientific Reports} {\bf 5}, 13071 (2015).
 
\bibitem{Dawes1980}  Dawes, R. M.  Social dilemmas. {\em Annu. Rev. Psychol.} {\bf 31}, 169-193 (1980).

\bibitem{sasidevan} Sasidevan, V. \&  Dhar, D.  Strategy switches and
co-action equilibria in a minority game. {\em Physica A} {\bf 402}, 306-317 (2014).

\bibitem{fundenberg} Fundenberg, D. \& Maskin, E.  The folk theorem in repeated games with discounting or with incomplete information. {\em Econometrica} {\bf 54(3)}, 533-554 (1986).

\bibitem{nowak} Nowak, M. \&  Sigmund, K.  A strategy of win-stay, lose-shift that outperforms tit-for-tat in the Prisoner's Dilemma game. {\em Nature} {\bf 364(6432)}, 56-58 (1993).

\bibitem{Kraines1989} Kraines, D. \& Kraines, V. Pavlov and the
Prisoner's dilemma. {\em Theor. Decis.} {\bf 26}, 47 - 49 (1989).

\bibitem{Imhof2007} Imhof, L. A., Fudenberg, D. \&  Nowak M. A.
Tit-for-tat or win-stay, lose-shift? {\em J. Theor.
Biol.} {\bf 247}, 574 - 580 (2007). 

\bibitem{domjan}  Domjan, M. \&  Burkhardl, B.  {\em The principles of learning and behaviour} (Brooks/Cole, Monterey, 1986).

\bibitem{wedekind}  Wedekind, C. \& Milinski, M.  Human cooperation in the simultaneous and the alternating prisoner's dilemma: Pavlov versus generous tit-for-tat. {\em Proc. Natl. Acad. Sci. USA} {\bf 93(7)}, 2686-2689 (1996).

\bibitem{nowak_2} Nowak, M. A. \& May, R. M.  Evolutionary games and spatial chaos. {\em Nature} {\bf 359(6398)}, 826-829 (1992).

\bibitem{Davis1993} Davis, D. D. \& Holt, C. A. {\em Experimental
Economics} (Princeton University Press, Princeton, 1993)

\bibitem{Olson1965} Olson, M. {\em The Logic of Collective Action}
(Harvard University Press, Cambridge, 1965).

\bibitem{Hauert2003} Hauert, C. \& Szabo, G. Prisoner's  dilemma  and
public  goods  games in  different  geometries:  Compulsory  versus
voluntary  interactions. {\em Complexity} {\bf 8(4)}, 31-38 (2003).

\bibitem{Kollock1998}  Kollock, P.   Social dilemmas: The anatomy of
cooperation. {\em Annu. Rev. Sociol.} {\bf 24}, 183-214 (1998)

\bibitem{hauert} Hauert, C. H. \&  Schuster, H. G.  Effects of
increasing the number of players and memory size in the iterated
prisoner's dilemma: A numerical approach. {\em  P.
Roy. Soc. Lond. B} {\bf 264(1381)}, 513-519 (1997).

\bibitem{boyd} Boyd, R. \& Richerson, P. J.  The evolution of
reciprocity in sizable groups. {\em J. Theor. Biol.} {\bf 132}, 337-356 (1988).

\bibitem{axelrod_2} Axelrod, R. \& Dion, D.  The further evolution of cooperation. {\em Science}  {\bf 242(4884)}, 1385-1390 (1988).


\bibitem{Sally1995} Sally, D.  Conversation and cooperation in
social dilemmas: A meta-analysis of experiments from 1958 to 1992. {\em
Ration. Soc.} {\bf 7(1)}, 58-92 (1995).

\bibitem{Nowak2006} Nowak, M. A.  Five rules for the evolution of cooperation. {\em Science}  {\bf 314(5805)}, 1560-1563 (2006).

\bibitem{Hamilton1964} Hamilton, W. D.  The genetical evolution of
social behaviour. II. {\em J. Theor. Biol.} {\bf 7(1)}, 17-52 (1964).

\bibitem{Nowak1998} Nowak, M. A. \&  Sigmund, K.  Evolution of indirect reciprocity by image scoring. {\em Nature} {\bf 393(6685)}, 573-577 (1998).

\bibitem{Lieberman2005} Lieberman, E., Hauert, C. \&  Nowak, M. A.  Evolutionary dynamics on graphs. {\em Nature} {\bf 433(7023)}, 312-316 (2005).

\bibitem{Traulsen2006} Traulsen, A. \& Nowak, M. A.  Evolution of cooperation by multilevel selection. {\em Proc. Natl. Acad. Sci. USA} {\bf103(29)}, 10952-10955 (2006).

\bibitem{radner}  Radner, R.  Can bounded rationality resolve the
prisoner’s dilemma?  in {\em Essays in honor of Gerard Debreu} (eds
Mas-Colell, A. \& Hildenbrand, W.) Ch. 20, 387-399 (North-Holland, Amsterdam,
1986).

\bibitem{kreps} Kreps, D., Milgrom, P.,  Roberts, J. \&  Wilson, R.
Rational cooperation in the finitely repeated prisoners' dilemma. {\em
J. Econ. Theory} {\bf 27}, 245-252 (1982).

\bibitem{neyman_2} Neyman, A.  Bounded complexity justifies
cooperation in the finitely repeated prisoners' dilemma. {\em Econ.
Lett.} {\bf 19(3)}, 227-229 (1985).

\bibitem{pettit} Pettit, P. \& Sugden, R.  The backward induction
paradox. {\em J. Philos.} {\bf 86(4)}, 169-182 (1989).

\bibitem{neyman_1} Neyman, A.  Cooperation in repeated games when the number of stages is not commonly known.  {\em Econometrica} {\bf 67(1)}, 45-64 (1999). 






\bibitem{Perc2008} Perc, M. \&  Szolnoki, A. Social diversity and promotion of cooperation in the spatial prisoner’s dilemma game. {\em Phys. Rev. E} {\bf 77}, 011904 (2008).




\bibitem{Szolnoki2012} Szolnoki, A. \&  Perc, M. Conditional strategies and the evolution of cooperation in spatial public goods games. {\em Phys. Rev. E} {\bf 85}, 026104 (2012).


\bibitem{Szolnoki2012b} Szolnoki, A.,  Perc, M. \&  Szabo, G. Defense
mechanisms of empathetic players in the spatial ultimatum game. {\em
Phys. Rev. Lett.} {\bf 109}, 078701 (2012).

\bibitem{Szolnoki2016} Szolnoki, A. \& Perc, M. Collective influence in
evolutionary social dilemmas. {\em EPL} {\bf 113}, 58004 (2016).










\bibitem{Vukov2006} Vukov, J.,  Szabo, G. \&  Szolnoki, A. Cooperation in the noisy case: Prisoner’s dilemma game on two types of regular random graphs. {\em Phys. Rev. E} {\bf 73}, 067103 (2006).

\bibitem{Szolnoki2009}  Szolnoki, A., Vukov, J. \& Szabo, G. Selection of noise level in strategy adoption for spatial social dilemmas. {\em Phys. Rev. E} {\bf 80}, 056112 (2009).

\bibitem{Neill2001} Neill, D. B. Optimality under noise: Higher memory
strategies for the alternating Prisoner's Dilemma. {\em J.
Theor. Biol.} {\bf 211}, 159 - 180 (2001).

\bibitem{Posch1999} Posch, M. Win-stay, lose-shift strategies for
repeated games - Memory length, aspiration levels and noise. {\em J.
Theor. Biol.}, {\bf 198}, 183 - 195 (1999).


\bibitem{Andreoni1995} Andreoni, J.  Cooperation in public-goods
experiments: kindness or confusion? {\em Am. Econ. Rev.} {\bf 85(4)}, 891-904 (1995)
\bibitem{Ledyard1997}  Ledyard, J. O.  Public goods: A survey of
experimental research, in {\em The Handbook of Experimental
Economics} (eds. Kagel, J. H. \& Roth, A. E.) Ch. 2, 111-193 (Princeton Univ. Press,
Princeton, NJ, 1997).


\bibitem{Riolo2001} Riolo, R. L., Cohen, M. D. \& Axelrod, R. Evolution
of cooperation without reciprocity. {\em Nature} {\bf 414(6862)}, 441 - 443 (2001).



\end{thebibliography}
\end{document}